\documentclass[11pt]{article}
\usepackage[margin=1in]{geometry}
\usepackage[utf8]{inputenc}
\usepackage[T1]{fontenc}
\usepackage{lmodern}
\usepackage{authblk}
\usepackage{graphicx}
\usepackage{bm}
\usepackage{xspace}
\usepackage{color}
\usepackage{comment}
\usepackage{amsmath,amssymb}
\usepackage[numbers,sort&compress]{natbib}
\usepackage{url}
\usepackage[hidelinks]{hyperref}
\title{Beyond the Flat-Band Paradigm: Open Orbits as a Route to Correlated States in Dispersive Moir\'e Minibands}
\author[1,2]{Zihao Wang}
\author[3,4]{Pablo M. Perez-Piskunow}
\author[5,6]{Calvin Pei Yu Wong}
\author[7]{Matthew Holwill}
\author[1]{Hao Sun}
\author[5,6]{Wei Fu}
\author[8]{Junxiong Hu}
\author[8]{Zheng Yuntian}
\author[1]{Zhang Xuanrui}
\author[9]{T. Taniguchi}
\author[9]{K. Watanabe}
\author[8]{A. Ariando}
\author[10]{Geliang Yu}
\author[1]{Li Lin}
\author[11,5,6]{Kuan Eng Johnson Goh}
\author[3,4]{Stephan Roche}
\author[12,13]{Jeil Jung}
\author[1,2]{Kostya S. Novoselov}
\author[12]{Nicolas Leconte}
\affil[1]{Institute for Functional Intelligent Materials, National University of Singapore, Singapore 117544}
\affil[2]{Department of Materials Science and Engineering, National University of Singapore, Singapore 117575}
\affil[3]{Catalan Institute of Nanoscience and Nanotechnology (ICN2), CSIC and BIST, Campus UAB, Bellaterra (Barcelona) 08193, Spain}
\affil[4]{ICREA--Instituci\'o Catalana de Recerca i Estudis Avan\c{c}ats, Barcelona 08010, Spain}
\affil[5]{Institute of Materials Research \& Engineering (IMRE), Agency for Science, Technology and Research (A*STAR), 2 Fusionopolis Way, \#08-03 Innovis, Singapore 138634, Republic of Singapore}
\affil[6]{Centre for Quantum Technologies, National University of Singapore, 3 Science Drive 2, Singapore 117543, Republic of Singapore}
\affil[7]{Department of Physics and Astronomy, University of Manchester, Manchester M13 9PL, United Kingdom}
\affil[8]{Department of Physics, National University of Singapore, Singapore 117544}
\affil[9]{National Institute for Materials Science, Japan}
\affil[10]{National Laboratory of Solid-State Microstructures, School of Physics, Nanjing University, Nanjing 210093, China}
\affil[11]{Temasek Laboratories@Nanyang Technological University, 50 Nanyang Drive, Research Techno Plaza, BorderX Block, 9th Storey, Singapore 637553, Singapore}
\affil[12]{Department of Physics, University of Seoul, Seoul 02504, Korea}
\affil[13]{Department of Smart Cities, University of Seoul, Seoul 02504, Korea}
\date{}
\begin{document}
\maketitle
\begin{center}\small\textit{Correspondence:} zihao.wang@nus.edu.sg; kejgoh@yahoo.com; jeiljung@uos.ac.kr; kostya@nus.edu.sg; nicolas.leconte@icn2.cat\end{center}
\begin{abstract}
The realization of fractional Chern insulators has predominantly relied on the engineering of isolated flat bands in complex bilayer or multilayer moir\'e systems with ideal quantum geometry. Whether such strongly correlated topological phases can emerge in theoretically more tractable systems with genuinely dispersive energy bands, and by what mechanism, remains a fundamental open question. Here, we propose a mechanism for magnetic-field-induced strong correlations in the highly dispersive bands of a monolayer graphene--hexagonal boron nitride superlattice. Operating in the low-field classical-to-quantum crossover regime of the second moir\'e miniband, we demonstrate that the interplay between the moir\'e potential and weak magnetic fields ($B<1.5$ T) profoundly alters electron dynamics. Saddle-point van Hove singularities give rise to extended open orbits that strongly suppress carrier propagation in magnetic fields, potentially favoring a regime in which Coulomb interactions become comparatively more important, while the weak magnetic field lifts valley degeneracy and trigonal warping redistributes Berry curvature. Through high-resolution magnetotransport and temperature-dependent measurements, we observe fractional-slope transport features together with signatures of interaction-driven incompressibility in this dispersive-band system. Because the Hall conductivity is not fully quantized and the longitudinal resistance remains finite, we refer to this phenomenology as \textit{incipient FCI} behavior: transport signatures consistent with the developing fractional Chern insulating state, without claiming a fully developed FCI phase established by complete Hall quantization and vanishing longitudinal resistance. Our results  support a possible fermiology-driven route  toward correlated topological  states in  dispersive bands in the classical-to-quantum crossover regime, distinct from  the conventional flat-band  scenario.
\end{abstract}

\section*{Introduction}
The fractional Chern insulator (FCI), a lattice analogue of the fractional quantum Hall effect~\cite{PhysRevX.1.021014,LIU2024515}, has largely been understood within a flat-band paradigm where the quenching of carrier kinetic energy amplifies electron-electron interactions. 
Within this flat-band framework, experimental evidence for FCIs has been reported in twisted bilayer graphene~\cite{Xie2021}, Bernal bilayer graphene/hBN Hofstadter bands~\cite{Spanton2018}, twisted MoTe$_2$~\cite{Cai2023-ac,Park2023-vf,Zeng2023-ik,PhysRevX.13.031037}, and rhombohedral multilayer graphene~\cite{Lu2024}. Related integer Chern-insulator phases have also been observed in rhombohedral graphene~\cite{Sha2024,Han2023-cz,Wang2024}. Theoretical studies describe FCI formation in single-particle moir\'e flat bands~\cite{PhysRevResearch.2.023237,PhysRevLett.132.036501,PhysRevLett.132.096602} and propose interaction-driven topological flat bands in rhombohedral graphene~\cite{PhysRevLett.133.206502,PhysRevLett.133.206504,PhysRevLett.133.206503,PhysRevB.110.075109,arxiv2312.11617}. 

A key open question is whether such strict flat-band conditions are necessary. Recent
theoretical advances have begun to challenge the conventional criteria for FCI formation, with
predictions that fractional Chern physics can persist far from the ideal quantum-geometry limit~\cite{Oreg,Chang,2510.15027,Lin2026} and systematic searches identifying unconventional regimes beyond Landau-level mimicry~\cite{Fonseca}. 
More broadly, saddle points and the associated van Hove singularities have long been recognized as a mechanism for enhancing interaction effects in otherwise dispersive bands, including proposals for unconventional superconductivity ~\cite{Markiewicz1994,PhysRevLett.122.026801} in genuinely
dispersive band structures. Experimentally isolating its effect, however, remains largely unexplored in systems such as twisted bilayer graphene, where flat bands and van Hove singularities typically
coexist near the magic angle.

Previous studies of graphene/hBN have established its miniband structure, relaxation effects, and energy gaps~\cite{woods,Wang:2016ii,Yankowitz:2014bv,Wallbank:2013ep,Chen2014,PhysRevLett.111.266801,RibeiroPalau2018,10.1021/acs.nanolett.8b03423}, while magnetotransport studies have revealed Hofstadter spectra and quantum Hall effects at higher fields~\cite{Dean2013,Ponomarenko2013,Hunt:2013ef}. In contrast, the low-field classical-to-quantized crossover ($0.5$--$1.5$ T)~\cite{Bocarsly2024} remains comparatively unexplored. Here, the magnetic length
still exceeds the moir\'e period, so electron dynamics are not yet governed by a fully developed
Landau or Hofstadter spectrum. Instead, they are strongly shaped by the detailed fermiology of the
dispersive minibands, including trigonal warping, pocket multiplicity, and open semiclassical
orbits. While negligible near charge neutrality, these effects become highly pronounced near the
secondary Dirac point (SDP) due to strong moiré-driven Fermi surface reconstruction. In this region, low-field magnetotransport provides a highly sensitive probe of moiré miniband
fermiology, revealing emergent physics inaccessible at both zero field and the high-field
Hofstadter limit.

Here, we propose and experimentally support an alternative pathway to correlated topological states that combines dispersive fermiology with non-quantized weak magnetic fields. Throughout this work, we use the term \textit{incipient FCI} behavior to describe fractional transport signatures consistent with a developing fractional Chern insulating state, while explicitly distinguishing them from a fully developed FCI with quantized Hall conductivity and vanishing longitudinal resistance. Our proposed mechanism for this incipient behavior relies on the interplay of the following three key ingredients.

\begin{itemize}

\item \textbf{Open-orbit-induced suppression of carrier propagation:} Crucially, saddle points in the miniband give rise to open orbits that traverse the Brillouin zone. At finite magnetic fields, these extended trajectories, rooted in the locally flat dispersion near the saddle points, strongly suppress carrier propagation and promote repeated momentum reversals and scattering near the zone boundaries, thereby increasing the relative importance of electron-electron interactions prior to full Landau quantization.

\item \textbf{Trigonal-warping-induced concentration of topological weight:} The moir\'e potential generates zero-field minibands that remain dispersive yet carry nonzero valley Chern numbers, while trigonal warping concentrates their topological weight in specific regions of the Brillouin zone. The resulting Fermi-pocket evolution organizes the integer transport background and identifies the momentum-space sector relevant to the emergent correlated state.

\item \textbf{Weak magnetic-field-induced valley splitting:} A weak magnetic field lifts the valley degeneracy and reorganizes these minibands into precursors of valley-split anomalous Landau levels before full Landau quantization sets in, revealing the topology of an otherwise valley-cancelled Chern miniband in the single-particle picture.

\end{itemize}

Our transport measurements in monolayer graphene--hBN superlattices support this mechanism. In
this system, the single-particle minibands remain valley degenerate and strongly dispersive, and
are unlikely to become isolated through interaction effects alone~\cite{Jung:1jb}, making it an ideal testbed for
the proposed mechanism. We first establish the persistence of open orbits up to $~\sim 1.5$ T through
large magnetoresistance transport measurements, real-space transport simulations, and field-
dependent current-density maps, confirming that the relevant regime is reached prior to full
Landau quantization. Near one such open orbit, we then observe fractional transport signatures consistent with incipient FCI behavior at $B<1.5$ T, identified by their characteristic slopes in the
$n$--$B$ plane. Along these fractional trajectories, the longitudinal-resistance minima deepen and the Hall response develops plateau-like features upon cooling, 
consistent with the fragile many-body nature of the fractional correlations. These fractional features
emerge when the three aforementioned conditions coexist, appearing in a narrow
density range associated with the second miniband and accompanied by a complex integer
background arising from trigonal-warping-driven pocket evolution. Together, these results suggest a new route toward fractional topological phenomena in genuinely dispersive moiré minibands.

\section*{Global magnetotransport landscape and identification of the low-field regime}

\begin{figure}%
\centering

\includegraphics[width=14.1cm]{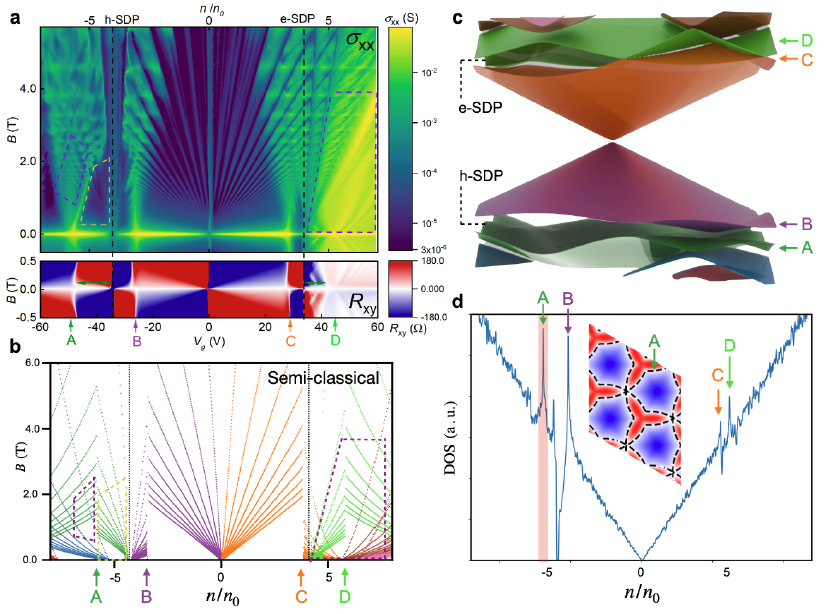}
\caption{\textbf{Full spectrum magnetic-field dependence of electronic states in G-hBN superlattice and signatures of incipient FCI behavior.}
\textbf{a} Displays $\sigma_\text{xx}$ and $R_\text{xy}$, highlighting hole and electron SDP (black dashed lines), Landau level splitting (yellow-dashed area), and non-linear Wannier diagram region (purple dashed area). vHS positions are marked as \textbf{A} to \textbf{D}.
\textbf{b} Semi-classical LL predictions based on zero-field Fermi contours, with colors indicating specific minibands. \textbf{c} shows the minibands corresponding to the Landau levels.
\textbf{d} highlights the DOS vHS caused by the open orbits around saddle points, and its inset shows the open orbit saddle point for \textbf{A} as a dashed line, separating the hole-type carriers in red and the electron-type carriers in blue.
}
\label{fig1}
\end{figure}

Figure~\ref{fig1} establishes a global perspective of the graphene--hBN superlattice with a 14.5 nm moiré period. We present the full magnetotransport mapping (Fig. 1\textbf{a}), alongside semi-classical Landau level predictions (Fig. 1\textbf{b}), the calculated band structure (Fig. 1\textbf{c}), and the corresponding density of states (Fig. 1\textbf{d}). The primary objective of this global overview is to locate the relevant electronic regimes within the full magnetotransport spectrum, explicitly separating the low-field miniband physics from the high-field Hofstadter response. In particular, the low-field regime reveals two phenomena that would otherwise be obscured at high fields: open orbits emerging at the saddle-point energies, and potentially interesting fractional features sustained by the zero-field miniband fermiology that remains active before full Landau quantization sets in.

The magnetoresistance map in panel \textbf{a} reveals the familiar Hofstadter butterfly patterns in the high-field regime, where magnetic quantization reorganizes the minibands into Hofstadter subbands whose structure is dominated by Landau-level physics. However, our primary focus lies in the low-field regime ($B < 1.5$ T). Here, the magnetic length, defined as $l_{B} = \sqrt{\hbar / eB}$
is larger than $20$ nm. Given that the moir\'e period of this sample is only $14.5$ nm (estimated from the carrier density at the SDP), the momentum-dependent moir\'e band structure remains the primary driver of the electronic properties in this regime, rather than the nearly dispersionless Landau-level spectrum that dominates at higher magnetic fields. To resolve the fine features in this regime with high energy resolution, we perform measurements at ultralow temperatures down to $10$mK, using a $2$ nA excitation current, and utilizing a Second Order Low Pass Filter (details in the Supplementary Information). 
Within this weak-field regime, van Hove singularities associated with miniband saddle points are identified through sign reversals in $R_\text{xy}$ and corresponding dips in $\sigma_\text{xx}$, labelled A-D in Fig. 1\textbf{a}. Their positions agree with the saddle points in the calculated band structure in Fig. 1\textbf{c} and the DOS peaks in Fig. 1\textbf{d}. These DOS peaks arise from the divergent density of states at each saddle point. At weak magnetic fields, the corresponding Fermi contours evolve into open semiclassical orbits, where electron-like and hole-like segments meet near the saddle-point energies. 
The dark green double arrows in the $R_\text{xy}$ map highlight a pronounced hole-electron asymmetry in the region between vHS and SDP. This experimental manifestation is matched with the band structure in panel \textbf{c}, which shows that beyond the SDP, miniband overlap is more extensive on the electron side, whereas the hole side exhibits considerably less overlap, giving rise to a single-band region enclosed by the yellow-dashed lines.  

There, we will locate the three ingredients of our proposed mechanism governing correlated states in dispersive bands and present high-resolution transport data revealing signatures of incipient FCI behavior in the final Section.

To simulate our experimental Landau level Wannier diagrams,
we develop a semi-classical approach (Fig.~\ref{fig1}\textbf{b}),
tailored to the near-zero-field regime where full quantum simulations are computationally intractable for realistic sample sizes. Specifically, our novel semi-classical method uses a tailored implementation of a Marching Square algorithm to extract constant-energy Fermi contours from each miniband. The corresponding contour areas are then used to estimate the Landau-level spectrum semiclassically. 
For methodological details, including the periodic-contour unwrapping procedure and a direct benchmark of the semiclassical construction against full quantum calculations in both a toy model and the realistic graphene--hBN miniband structure, see Figs.~S5 and S6 of the Supplementary Information. A three-dimensional view of the calculated minibands is additionally provided in Fig.~S7. Overall, our semi-classical simulation in panel \textbf{b} reproduces the main single-particle features in panel \textbf{a}, with corresponding saddle points and regions of interest coded by the same color. Importantly, each saddle point marks the transition between electron-like and hole-like Landau-level branches with opposite slopes within a same miniband in the Wannier diagram, reflecting the underlying Lifshitz transition in the zero-field Fermi surface.  

The remainder of the paper establishes the three ingredients of our mechanism in sequence. We first show that saddle-point open orbits survive deep into the low-field regime and strongly suppress carrier propagation. We then show that trigonal-warping-driven Fermi-pocket evolution organizes the integer background while concentrating the topological weight of the moir\'e-induced valley Chern miniband within a specific region of the moir\'e Brillouin zone. Next, we demonstrate that a weak magnetic field lifts the valley degeneracy in this dispersive Chern miniband. Having established these ingredients, we finally return to high-resolution transport and identify signatures of incipient FCI behavior precisely in the regime where all three conditions coexist.

\section*{Open orbits and suppressed carrier propagation at weak magnetic fields}

\begin{figure}[tbhp]%
\centering

\includegraphics[width=12.9cm]{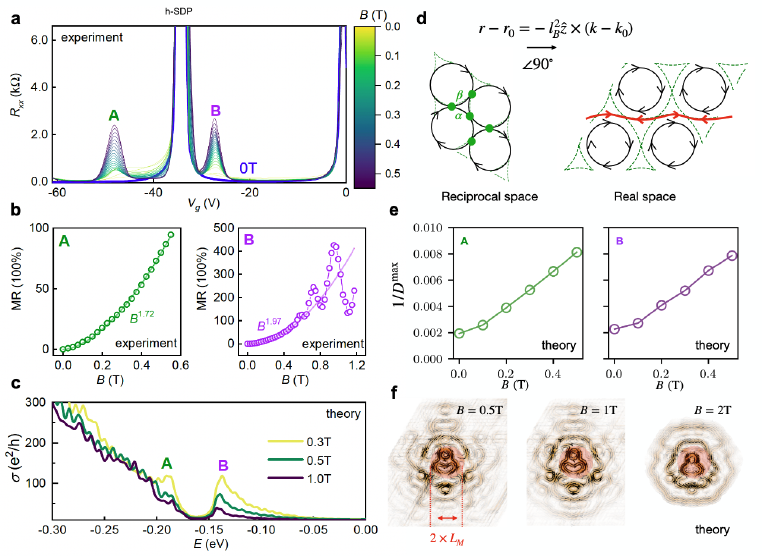}
\caption{\textbf{Open-orbit induced XMR.}
\textbf{a} Low-field $R_\text{xx}$ demonstrates positive XMR at hole-side singularities \textbf{A} and \textbf{B}, aligned with open orbits in minibands. \textbf{b} The hollow circles represent peak values at different magnetic fields, while the light-colored lines indicate the fitting curves. The magnetoresistance increases by several hundred-fold at low magnetic fields, indicating strongly suppressed carrier propagation at open-orbit energies \textbf{A} and \textbf{B}. \textbf{c} 
Theoretical estimates of the conductivity on the hole side. The highlighted regions near open orbits \textbf{A} and \textbf{B} show suppressed conductivity associated with open-orbit dynamics.
\textbf{d} correlates electron orbits in real space with k-space counterparts, following Onsager's approach~\cite{onsager1952}, with a $\angle 90^\circ$-rotation and scaling factor indicated in the figure.
Green dots mark large-angle scattering hot-spots. Closed (black) and extended (red) orbits in real-space sketch illustrate state-direction reversal along the red line, matching the resistivity increase, due to the increased scattering at the BZ edge. \textbf{e} Inverse of the saturation value of diffusion coefficient calculation, matching the MR trends in \textbf{b}.
\textbf{f} Local current density maps show extended open contours persisting up to $~0.5-1$T, and closed cyclotron orbits at $2$T. Red-shaded area indicates the wavefunction injection site.
}
\label{fig2}
\end{figure}

We next establish the first ingredient of the proposed mechanism: saddle-point open orbits that strongly suppress carrier propagation at weak magnetic fields. In Fig.~\ref{fig1}\textbf{d}, we identify a series of van Hove singularities (\textbf{A-D}) in the density of states, which are often discussed in terms of their enhancement of interactions~\cite{PhysRevB.107.184504, Markiewicz1994,PhysRevLett.122.026801}. Here, we extend this traditional picture by also considering their finite-field transport consequences. Each vHS corresponds to a saddle point separating hole-like and electron-like carriers, as evidenced by the sign changes in $R_{xy}$ shown in the bottom panel of Fig.~\ref{fig1}\textbf{a}. In momentum space, this saddle point marks a Lifshitz transition between closed hole-like and electron-like Fermi contours. At the transition energy, the constant-energy contour becomes open and traverses the moir\'e Brillouin zone, giving rise to the open orbit illustrated by the dashed curve in the inset of Fig.~\ref{fig1}\textbf{d}. Such open semiclassical trajectories strongly modify carrier transport in a magnetic field because electrons drift over extended distances in one momentum-space direction instead of executing closed cyclotron motion.

To investigate this transport effect of open orbits on electron transport under low magnetic fields, where the candidate fractional features are observed, we measure the $R_\text{xx}$ resistance at low magnetic fields before the onset of SdH oscillations.

As Fig.~2\textbf{a} shows, $R_{xx}$ at $B = 0$~T (blue curve) exhibits no peaks at saddle points \textbf{A} and \textbf{B}~\cite{Markiewicz1994}. With increasing magnetic field, however, prominent resistance peaks emerge at these saddle points, revealing an extremely large positive magnetoresistance (XMR). Strikingly, this behaviour is opposite to that of the surrounding background, where $R_{xx}$ instead decreases with field, namely a negative magnetoresistance consistent with weak localization. This sign contrast between the saddle points and the background underscores the distinct transport mechanism operating at \textbf{A} and \textbf{B}. Furthermore, Fig.\ref{fig2}\textbf{b} demonstrates the positive XMR with a near-quadratic fit. At $0.55$T, XMR values reach $10000\%$ for both vHS \textbf{A} and \textbf{B}.

Above 0.55 T, incipient SdH oscillations gradually emerge, as indicated by the purple open circles, consistent with the expected crossover toward Landau quantization. The appearance of XMR at sharply defined carrier densities points toward an intrinsic band-topological origin, while the near-quadratic $B$ dependence at low fields is consistent with open-orbit transport~\cite{PhysRev.138.A88,PhysRevB.23.6294,PhysRev.175.813,PhysRevB.104.115104,PhysRevLett.111.056601,PhysRevResearch.2.022042,PhysRevB.99.035142,Markiewicz1994}. Indeed, in the near-quadratic open-orbit regime $\rho_{xx}$ scales as $B^{q}$ with $q=2$ at leading order; subleading contributions can give effective exponents below two, consistent with the experimentally observed exponents in the range $q = 1.72\text{--}1.97$  (see Supplementary Information).

Intuitively, the hundredfold increase in resistance indicates that carrier propagation is strongly suppressed,  which  we propose may favor a regime where interactions become comparatively more important. This transport suppression should not be interpreted as a flattening of the zero-field miniband or a reduction of its microscopic bandwidth: the band remains dispersive, while the field-driven open-orbit dynamics reduce net diffusion through repeated momentum reversals and scattering. 

Moreover, electron--hole compensation is unlikely to account for the XMR at saddle points \textbf{A}--\textbf{D}. It is absent in the calculated band structure at saddle points \textbf{B} and \textbf{C}, where electron and hole pockets do not coexist, and where band overlap occurs the positive magnetoresistance remains sharply confined to the saddle-point densities rather than extending over the broader density range expected for near compensation (see Supplementary Information).

To understand microscopically how open orbits achieve this suppression, we go beyond the reciprocal-space picture and examine the real-space dynamics. 
The usual reciprocal-space argument, that kinetic energy, whether $E_{\mathrm{kin}} = \hbar^{2}|\mathbf{k}|^{2}/(2m^{*})$ or $\hbar v_{F} |\mathbf{k}|$, is unchanged under large-angle scattering ($\mathbf{k} \rightarrow -\mathbf{k}$), obscures the physical mechanism at play.
In real space, wavepacket reversals at saddle-point scattering hot spots correspond to back-and-forth motion, which not only reduces the net carrier drift but also increases the probability of repeated scattering from disorder, motivating the use of quantum transport simulations. In Fig.~\ref{fig2}\textbf{c}, We present real-space quantum simulations of wavepacket evolution in samples with low-concentration, short-range Gaussian disorder (see Supplementary Information for details). At energies near the open orbits, these simulations reveal a pronounced drop in conductivity with increasing magnetic field, whereas at energies away from the open orbits, conductivity remains nearly field-independent, in good agreement with the transport measurements in Fig.~\ref{fig2}\textbf{a}. The real-space origin of this energy-selective suppression is illustrated in panel \textbf{d}: at open-orbit energies, the frequent momentum reversals of a charge carrier (red trajectory) produce a highly non-directional real-space propagation, naturally explaining the increase in resistivity as the carriers can no longer propagate freely. The magnetic field enhances the scattering of carriers from one orbit to a nearby touching orbit, leading to momentum reversal even when the energy does not exactly match the open-orbit condition~\cite{Markiewicz1994}. This energy broadening of the open-orbit effect on transport coincides with the density range where the candidate fractional features are observed.
Finally, the inverse of the corresponding diffusion coefficients at saturation in panel \textbf{e} for energies \textbf{A} and \textbf{B} supports the picture of suppressed carrier diffusion with increasing magnetic field and qualitatively matches the trends from panel \textbf{b}. Such outcomes complement previously inferred reduction in Fermi velocity for the bulk states at saddle points~\cite{PhysRevLett.121.137701}, while also accounting for mesoscopic disorder effects that are ignored in the ballistic regime stemming from the band structure. 

We note that the open-orbit picture, including the influence of the saddle-point-enhanced DOS and the suppression of carrier propagation in the candidate fractional regime, relies on the assumption that Shubnikov-de Haas oscillations and Landau level quantization have not yet fully developed, so that carriers still feel the zero-field band structure. To confirm this assumption, in panel \textbf{f}, we illustrate the real-space current-density maps that show a transition from open orbit to closed orbits around $2T$, qualitatively agreeing with the onset of SdH oscillations in panel \textbf{b} for open orbit \textbf{B}. Additional field-resolved current-density maps between $0.1$ and $0.9$ T, presented in Fig.~S10, resolve the gradual formation of closed trajectories and the transient coexistence of orbit sizes across this crossover. 

Together, the measurements and quantum transport simulations establish the suppression of carrier propagation, providing a single-particle basis for investigating the proposed enhancement of interaction effects. We next turn to the second ingredient of our mechanism, the detailed Fermi-pocket evolution near the same saddle-point region, and show that the observed integer sequences arise naturally from trigonal-warped single-particle fermiology.

\section*{Single-particle Band Fermiology Shaped by Trigonal Warping}

\begin{figure}%
\centering

\includegraphics[width=15.2cm]{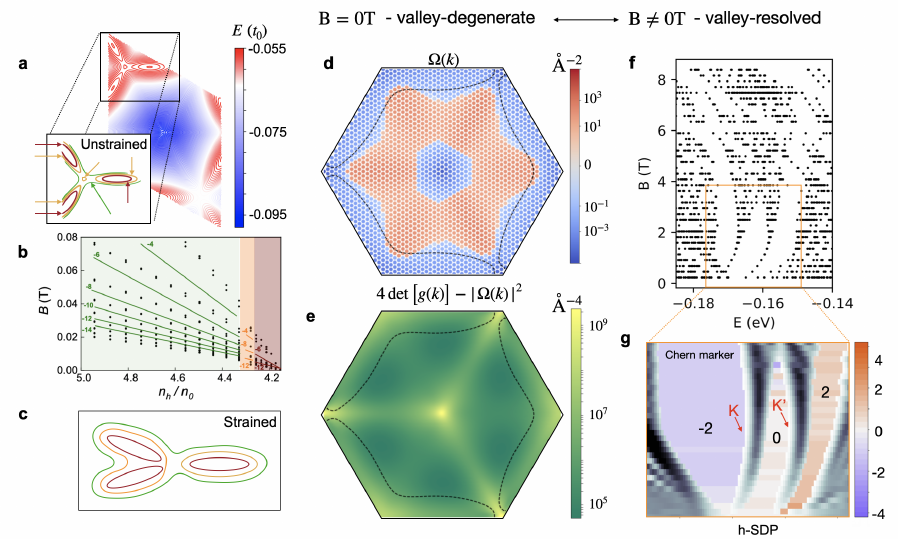}

\caption{\textbf{Single-particle fermiology and topological structure of the active miniband.} \textbf{a} Fermi contours showing the trigonal-warping induced merging of Fermi pockets as highlighted in the inset. \textbf{b} Semi-classical confirmation of splitting using the contours from \textbf{a} where we plot a subset of LLs for each energy. Same-color arrows are used to correlate the pockets in \textbf{a} with their corresponding LL behavior. Diophantine equation fitting of the gaps separating the semi-classical Landau levels. %
The values give the slopes $t$ following the experimental sequences, as given in the text, but using $m'=2m$ instead of $m$, where the additional factor $2$ accounts for the valley degree of freedom, because our semi-classical approach based on the zero-field bands, by definition, does not lift the valley-degeneracy. \textbf{c} illustrates the 2-pocket configuration discussed in the text, illustrated schematically as a possible strain-based interpretation of the broad experimental $-4m$ sequence.
\textbf{d} Berry curvature of one of the valley-degenerate zero-field minibands with blue (red) corresponding to negative (positive) Berry curvature. The dashed line marks the boundary between the outer $k_\text{red}$ points that are used in the main text and the rest of the BZ. \textbf{e} Illustration of the determinant condition.
\textbf{f} Eigenvalues under magnetic field using exact diagonalization near h-SDP. \textbf{g} Chern marker~\cite{PhysRevResearch.2.013229} calculations to confirm the valley-degeneracy lifting in the one-pocket regime. 
}
\label{fig3}
\end{figure}

Having established that open orbits strongly modify transport near the saddle points, we now identify which features in the same low-field region can be accounted for without invoking e-e interactions. This step is essential because the candidate fractional features appear together with a set of integer lines; to isolate the fractional contribution, we must first establish which lines have a single-particle origin. We do so by analyzing the zero-field Fermi contours and their semiclassical Landau level sequences, which relate the main integer sequences to those expected from the trigonal-warped band structure.

In Fig.~\ref{fig3}\textbf{a-b}, we provide a close-up of the semi-classical LL integer filling level predictions that can be attributed to the specific band structure's influence on the electrons' SdH oscillations in the classical-to-quantized regime. As shown in Fig.~\ref{fig3}\textbf{a}, trigonal warping distorts the cone's circular Fermi contour into a multi-pocket structure for each valley, echoing the behaviour of the primary Dirac cone in Bernal bilayer graphene~\cite{Zhou2022-cl}.

As the inset shows, the Fermi contours reveal a sequence of transitions in the second valence miniband: starting with a 3-pocket configuration at lower energy levels near the hole-side SDP (dark red), transitioning to a 4-pocket region at higher energies (orange), and eventually merging into a single-pocket structure (green). 

This sequence predicts a hierarchy of integer sequences dictated by the pocket multiplicity:  a 3-pocket regime with $\nu=-6,-12,\dots$, a 1-pocket regime with spacing $\nu=-2,-4,\dots$, and, depending on details of pocket merging, an intermediate $-4m$ sequence. As we will show experimentally below (Fig.~\ref{fig4}), these sequences are indeed observed and support the fermiology-based interpretation. In Fig.~\ref{fig3}\textbf{b}, the semiclassical Landau level calculations confirm the link between the number of pockets and the multiplicity of the Landau levels. The Diophantine equation fits to the gaps separating the semiclassical levels are shown as solid lines. In the 1-pocket regime, a $t=-2m$ sequence occurs, where the factor of 2 accounts for spin degeneracy.
The 3-pocket phase corresponds to a
$t=-6m$ sequence, where each pocket contributes $t=-2$ to the total Chern number. 

To account for the experimental $-4m$ sequence observed between the 3-pocket and 1-pocket regimes, we consider two possible scenarios.
In the first, our semiclassical calculations predict a narrow 4-pocket regime. To match the experimental sequence, mixing of Landau levels and partial cancellation of Chern numbers between inner and outer pockets would be required (\textit{i.e.} $-2-2-2+2=-4$). However, this region is theoretically very narrow (orange region in panel \textbf{b}), whereas the experimental signatures extend over a much broader energy range. This discrepancy suggests that an alternative explanation is needed. 

In a second, explicitly speculative, scenario, the experimentally inferred $14.5$~nm moir\'e period is close to the nominal strain-free aligned value of about $14$~nm. This comparison alone does not establish reconstruction or broken threefold symmetry. A possible anisotropic strain component could break the threefold symmetry of the pocket distribution. If present, such anisotropy could cause two pockets to merge first before combining with the third, resulting in an intermediate 
$-4m$ sequence that corresponds to a $2$-pocket regime sandwiched between the $3$-pocket and $1$-pocket phases. This possibility is illustrated only schematically in panel \textbf{c}, where the same colour coding is used and orange denotes the 2-pocket configuration. The precise assignment of the intermediate pocket configuration is not essential to our central conclusion: trigonal warping reorganizes the integer sequence and identifies a region containing substantial topological weight within the experimentally relevant momentum-space sector.

The above analysis relies entirely on semiclassical single-particle physics. Because it starts from the zero-field Fermi contours, this approach provides a direct connection between the observed Landau-level sequences and the underlying band structure, making the physical origin of each integer sequence transparent. By contrast, full quantum calculations performed directly at finite magnetic field produce a considerably denser Landau-level spectrum in which the correspondence between individual features and the zero-field fermiology is less transparent. For completeness, a high-resolution finite-field KPM calculation is shown in Fig.~S11, where the proliferation of trigonal-warping-induced splittings is directly resolved. Within this framework, the semiclassical analysis  accounts for the main aspects of the integer sequences, while the broad intermediate regime remains unresolved. The fractional  features are not reproduced by this model, supporting an interaction-based interpretation.

Importantly, the trigonal-warping-driven pocket evolution does more than determine the integer Landau-level spectrum. As the Fermi pockets merge, it also redistributes the electronic states within the miniband, concentrating them into specific regions of momentum space. As we show below, the experimentally relevant one-pocket regime contains a disproportionately large fraction of the miniband Berry curvature and valley Chern character. Trigonal warping therefore plays a dual role: it not only organizes the semiclassical transport background through the evolution of the Fermi pockets, but also identifies the momentum-space sector that carries most of the miniband's topological weight. Before returning to the experimental fractional features, we therefore establish that this active miniband possesses the required topological character and that a weak magnetic field lifts the valley cancellation that would otherwise mask it.

\section*{Valley-split Chern bands and topological character of the dispersive miniband}

The single-band 1-pocket regime closest to the open orbit is part of the wider non-zero Chern number \textit{dispersive} miniband (colored dark green in Fig.~\ref{fig1}\textbf{c}). Conventionally, to argue for the FCI formation in the single-particle picture, one usually checks whether the zero-field band geometry resembles that of a Landau level supporting Laughlin states~\cite{PhysRevResearch.5.L032022}.
The usual flat-band criteria favoring such FCI states include~\cite{PhysRevResearch.2.023237} (i) narrow bands, (ii) a relatively uniform distribution of the Berry curvature, and (iii) an ideal quantum metric. The latter can be characterized by the \textit{determinant condition} in its weaker form and by the \textit{trace condition} in its stronger form. Vortexability formalizes the ability to insert vortices while remaining within the band subspace and does not require uniform Berry curvature~\cite{Ledwith2023Vortexability}. 
In the ideal lowest-Landau-level limit, the geometric conditions are satisfied, providing a natural setting for incompressible fractional states. At the low fields relevant to this work, however, the system sits in the crossover between Landau level quantization and zero-field band physics, so this ideal limit is not reached. 

We therefore proceed under the assumption that the relevant topology is already encoded in the zero-field miniband. Calculating its Chern number, we obtain $-1$ (or $1$, depending on the valley), consistent with previous predictions of moiré-induced finite Chern numbers in graphene~\cite{doi:10.1073/pnas.1424760112,PhysRevLett.120.026802, PhysRevB.98.125408}. Comparisons among three tight-binding parametrizations, including a rigid structure and two relaxed models, are shown explicitly in Fig.~S8. The rigid parametrization yields a vanishing Chern number, whereas both relaxed parametrizations yield a finite valley Chern number. The agreement between the two relaxed descriptions shows that the existence of the finite Chern number is robust with respect to the choice between these relaxation models, while its absence in the rigid model shows that this conclusion depends on the structural parametrization. This relaxation modifies the moir\'e potential and band topology without requiring broken threefold symmetry and is therefore distinct from the possible anisotropic strain invoked above as a speculative explanation of the intermediate two-pocket regime. The parameters are taken from the literature without fitting, using a lattice mismatch of $1.8\%$ (see Supporting Information). We emphasize that this topology is intrinsic to the zero-field miniband and is not generated by the applied magnetic field.

We next calculate the Berry curvature [Fig.~\ref{fig3}\textbf{d}] and evaluate the determinant condition [Fig.~\ref{fig3}\textbf{e}] to assess the quantum geometry of the miniband. Although these quantum-geometric criteria are motivated by lowest-Landau-level physics, we use them here only as heuristic diagnostics of whether a restricted region of the dispersive miniband may provide favorable conditions for FCI states. This restricted momentum-space perspective is motivated by the trigonal-warping-driven reconstruction of the Fermi surface, which makes the low-field transport particularly sensitive to the sector adjacent to the open orbit. We therefore focus on the narrow range of hole carriers between the h-SDP and open orbit \textbf{A}, denoted $k_{\rm red}$. Remarkably, this region contains more than 80\% of the integrated Berry curvature associated with the total miniband Chern number, indicating that most of the miniband's topological weight is concentrated within the experimentally relevant momentum-space sector. To further characterize the Berry-curvature distribution in this active region, we calculate the normalized mean-standard deviation 
\begin{equation}
\sigma(\Omega(k_\text{red}))=\sqrt{\int_{BZ} \left(\frac{\Omega(k_\text{red})}{2\pi} - C\right)^2 dk_\text{red}},
\end{equation}
and obtain $\sigma[\Omega(k_{\rm red})]/\sigma[\Omega(k)]\sim0.5$. Within this heuristic comparison, the reduced Berry-curvature fluctuations suggest more favorable local quantum geometry in the active region than across the full miniband. 

Nevertheless, the determinant condition in Fig.~\ref{fig3}\textbf{e} still deviates substantially from its ideal value, showing that the ideal quantum-geometric conditions are not globally satisfied in this dispersive miniband. The stronger trace condition is consequently also not expected to hold globally.

Because the zero-field bands remain valley-degenerate, their opposite Chern numbers of $\pm1$ yield no net single-particle Chern response. As shown in Fig.~\ref{fig3}\textbf{f,g}, however, exact diagonalization and Chern-marker calculations~\cite{PhysRevResearch.2.013229}, performed within our tight-binding model with the magnetic field introduced through a Peierls substitution~\cite{PhysRevB.103.045402}, show that a finite magnetic field lifts this valley degeneracy~\cite{Moon:2014ke}. The calculated valley splitting is consistent with a field-induced orbital response~\cite{PhysRevB.81.195431}, providing a single-particle route toward valley-resolved occupation of the dispersive Chern minibands. Such magnetic-field-induced valley splitting is characteristic of anomalous Landau-level physics in graphene.

Taken together, Fig.~\ref{fig3} shows that the topology of the relevant valley Chern bands is already encoded in the zero-field dispersive miniband and remains relevant in the low-field crossover regime, before fully developed Landau quantization sets in. 

We have thus identified the three key ingredients of our proposed mechanism. Saddle-point open orbits strongly suppress carrier propagation, providing a regime in which interaction effects may become comparatively more important. Trigonal-warping-driven fermiology organizes the background integer sequences while concentrating the active Berry curvature within the experimentally relevant region of the Brillouin zone. Finally, the weak magnetic field lifts the valley degeneracy of the dispersive Chern miniband, providing a route toward valley-resolved occupation.

The remaining question is therefore whether fractional transport features emerge in the regime where these three ingredients coexist. We now return to the high-resolution transport data to address this question.

\section*{Experimental signatures of incipient FCI behavior}

\begin{figure}%
\centering

\includegraphics[width=12.6cm]{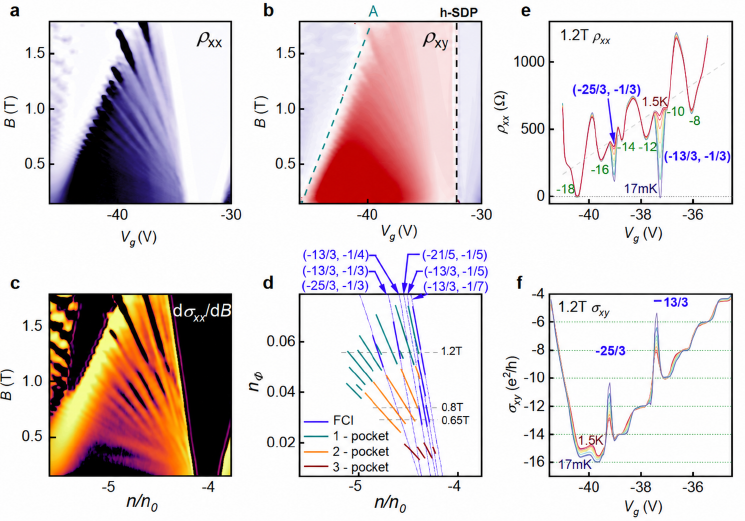}
\caption{\textbf{High-resolution transport signatures of incipient FCI behavior on the hole side, corresponding to the yellow-dashed region from Fig.~\ref{fig1}\textbf{a}.}
\textbf{a} $\rho_\text{xx}$ map.
\textbf{b} $\rho_\text{xy}$ map.
\textbf{c} $d\sigma_\text{xx}/dB$ representation enhancing the visibility of integer and fractional features.
\textbf{d} Schematic summary of the observed integer and fractional filling sequences. Dark green, orange, and dark red correspond respectively to the 1-pocket, 2-(or 4-)pocket, and 3-pocket regimes induced by trigonal warping. Blue lines mark the candidate fractional states. Panels~\textbf{e} and~\textbf{f} show temperature-dependent linecuts at $B=1.2$~T through the $(t,s)=(-13/3,-1/3)$ and $(-25/3,-1/3)$ features. The trajectories in panel~\textbf{d} are resolved over different magnetic-field ranges, so a single-field linecut does not show every candidate. The $(-21/5,-1/5)$ and $(-13/3,-1/5)$ features are shown in the $0.8$~T linecut in Fig.~S3; the temperature dependence presented here concerns only the two features at $1.2$~T.
\textbf{e} Temperature-dependent $\rho_\text{xx}$ linecuts at $B=1.2$T showing progressive deepening of the fractional minima associated with the $(t,s)=(-13/3,-1/3)$ and $(-25/3,-1/3)$ states as temperature decreases from $1.5$K to $17$mK.
\textbf{f} Corresponding $\sigma_{xy}$ linecuts showing the gradual evolution toward fractional Hall plateau-like behavior at the same fillings with decreasing temperature, consistent with the onset of interaction-driven incompressibility in the classical-to-quantized crossover regime. The integer states closely match the expected quantized $\sigma_{xy}$ values, while the small blue markers indicate the expected fractional $\sigma_{xy}$ values inferred from the slopes extracted from the $n$--$B$ mapping.
}
\label{fig4}
\end{figure}

To visualize the direct coexistence of trigonal-warping-governed single-particle fermiology and emergent
many-body correlations emerging near the open-orbit regime at weak magnetic fields, we focus on high-resolution
magnetotransport maps of the classical-to-quantized crossover regime. Figures~\ref{fig4}\textbf{a-d} present a
magnified view of the region enclosed by the yellow dashed lines in Fig.~\ref{fig1}\textbf{a}. Here, $\rho_{xx}$ and
$\rho_{xy}$ are displayed in panels~\textbf{a} and~\textbf{b}, respectively, while panel~\textbf{c} plots $d\sigma_{xx}/dB$ to
enhance the visibility of subtle transport features. Before turning to the fractional states, we first establish
the single-particle background. As summarized in panel~\textbf{d}, the integer states clearly separate into three
distinct regimes with different filling-factor spacings. These  are consistent with the  pocket-evolution interpretation discussed above, with the intermediate two-pocket assignment remaining speculative (labeled in green, orange, and dark red).

With this  interpretation of the integer background, we turn to the critical experimental test for our
proposed mechanism. As marked by the solid blue lines in panel~\textbf{d}, candidate fractional states appear
precisely within this same low-field window. This spatial co-location is essential because it occurs exactly
where the three ingredients of the mechanism coexist: the open-orbit-induced suppression of carrier
propagation, the trigonal-warping-driven organization of the integer background, and the weak-field
valley splitting of a dispersive Chern miniband. Crucially, because the surrounding
integer  sequences can largely be understood through single-particle Fermi-pocket evolution,  the additional fractional-slope features, which are not reproduced by our semiclassical  model, motivate an interpretation involving interactions.

To probe the nature of the fractional features,
Figs.~\ref{fig4}\textbf{e-f} present temperature-dependent linecuts at $B = 1.2\,\mathrm{T}$, cooling from $1.5\,
\mathrm{K}$ to $17\,\mathrm{mK}$. In panel~\textbf{e}, the $\rho_{xx}$ minima associated with the $(t,s) =
(-13/3,\,-1/3)$ and $(-25/3,\,-1/3)$ states progressively deepen as the temperature decreases in this temperature range, consistent
with the emergence of interaction-driven incompressibility. Panel~\textbf{f} shows the corresponding $
\sigma_{xy}$ linecuts in which the integer states closely follow the expected quantised $\sigma_{xy}$
values while the fractional states gradually evolve towards a fractional Hall plateau marked by a blue marker,
inferred from the slopes extracted from the $n$-$B$ map. In stark contrast to the temperature-insensitive integer background, the simultaneous temperature evolution of fractional features in both $\rho_{xx}$ and $\sigma_{xy}$ exhibits a
 pronounced temperature dependence  over the measured range. That indicates that the
fractional features are markedly more fragile upon warming.  Together with the fractional slopes, this thermal contrast  supports an interpretation in terms of many-body correlations developing within the low-field dispersive-miniband  regime, but does not establish their microscopic origin. These observations do not by themselves establish activated transport, and we do not extract or claim an activation gap.

The Hall conductivity does not reach the fractional values inferred from the fitted slopes, and $\rho_{xx}$ remains finite (Fig.~\ref{fig4}\textbf{f} and Fig.~S3). We therefore describe these observations as incipient FCI behavior. Possible contributions from residual dispersive states and the nonlinear Hall background are discussed tentatively in the Supplementary Information.

We now examine the fitted $t$ values in detail. A detailed analysis and quantitative verification of the fractional-state slopes are presented in Fig.~S4. First, the observed sequence of integer $t$ values in
Fig.~\ref{fig4}\textbf{d} is distinct from the Shubnikov-de Haas oscillations measured around the main Dirac
point: here, the integer states exhibit clear transitions between three distinct conductivity regimes (dark
red, orange, and green), accompanied by visible, systematic changes in the filling-factor spacing: $\Delta
t=6$ in the dark red regime ($-6, -12, -18, \dots$), $\Delta t=4$ in the orange regime ($-4, -8, -12, \dots$),
and $\Delta t=2$ in the green regime ($-8, -10, -12, -14, -16, \dots$). Superimposed on this background,
we identify fractional features corresponding to rational $(t,s)$ pairs (blue lines). Notably,  the fitted slope indices have denominators  $3$ or $5$, compatible with the odd-denominator hierarchy commonly associated with Abelian fractional quantum Hall states.~\cite{PhysRevLett.111.126802}. The unusually larger numerators in $t$ naturally
arise because the fractional states appear beyond the hole side SDP, i.e., in partially filled higher
minibands. The clearest signal corresponds to $t = -13/3 = -(4 + 1/3)$, which we tentatively associate with a fully occupied first hole-side miniband and a one-third-filled component of the second miniband. The other observed fractions, including $t=-21/5$ and $t=-25/3$, are less straightforward to assign to specific fillings of subsequent minibands. Further theoretical modeling will be required to fully rationalize the observed values of t in this
multiband regime.

\section*{Conclusions}

We have proposed a fermiology-driven route toward correlated topological states in dispersive moir\'e minibands under weak magnetic fields, and provided experimental evidence  consistent with this scenario in monolayer graphene aligned with hBN. The observed phenomenology emerges from the interplay of three ingredients: saddle-point open orbits that strongly suppress carrier propagation, trigonal-warping-driven pocket evolution that organizes the integer spectrum and reshapes the active momentum-space structure, and magnetic-field-induced valley splitting of dispersive Chern minibands. Within this framework, the observed fractional features arise  in the regime where these ingredients coexist, and are consistent with incipient FCI behavior prior to full Hofstadter quantization. Fundamentally, these findings motivate further investigation of correlated topological states beyond isolated flat bands, whether of single-particle origin or generated by interaction-driven mechanisms.

The role played by the magnetic field in our mechanism differs fundamentally from that in the Hofstadter regime, where strong magnetic fields dominate over the bare miniband dispersion and reorganize the spectrum into Hofstadter subbands or Landau levels that are often substantially flatter than the original bands. In our case, the magnetic field is too weak to fully quantize the spectrum, so the fractional features emerge from a regime in which the underlying dispersive miniband fermiology remains active. This mechanism is
likewise qualitatively distinct from the role of low magnetic fields in magic-angle twisted bilayer graphene, where the field acts on pre-existing flat bands~\cite{Xie2021}. Here, instead of generating flatness, the weak field acts on the existing fermiology: it helps suppress
carrier transport,  may enhance the relative importance of interactions, and exposes the
topology of otherwise valley-degenerate Chern minibands.

A further distinctive aspect of this work lies in the hitherto uncharted territory it addresses, namely high-energy superlattice minibands deep in the classical-to-quantized crossover regime and in the methodology developed to access it. Although graphene superlattices have been
studied extensively for more than a decade, this region has remained essentially
unexplored. The physics here extends far beyond the conventional evolution from Shubnikov-de Haas oscillations to the integer quantum Hall effect observed near charge neutrality. In addition to the signatures of incipient FCI behavior, we observe pocket evolution arising from trigonal warping, extremely large magnetoresistance associated with open orbits, and nonlinear Landau-level scaling in the Wannier diagram caused by overlapping minibands. To analyze these phenomena, we develop a semi-classical approach that avoids the prohibitive cost of fully quantum simulations at such low magnetic fields while faithfully reproducing the magnetotransport maps on both the hole and electron sides. Beyond the
present system, this provides a practical framework for separating semiclassical miniband features from additional low-field transport structures. 

The proposed mechanism  may extend to  other dispersive-band platforms. Extending this scenario requires more than the presence of a van Hove singularity or open orbits: the transport-active sector must carry substantial topological weight from a miniband with nonzero flavor-resolved Chern character. In graphene/hBN, inversion-symmetry breaking induced by hBN enables the valley Berry curvature. The relevant topological miniband should dominate low-energy transport, although it need not be globally isolated or flat. The interaction and flavor-splitting energy scales must exceed thermal and disorder broadening within the same energy and magnetic-field window, before conventional Landau quantization dominates the miniband structure. These considerations identify promising regimes in which to search for incipient FCI behavior. Possible platforms include large-angle twisted TMDs and Bernal bilayer or rhombohedral trilayer graphene, particularly when interfaced with substrates such as WSe$_2$ or MoS$_2$. More broadly, our results suggest that
weak-field dispersive regimes may offer an alternative route toward interaction-driven topological phenomena beyond the conventional flat-band paradigm. Looking ahead, the appearance of signatures of incipient FCI behavior at such low magnetic fields further suggests a possible pathway toward the fractional quantum anomalous Hall effect, although at zero field an additional mechanism, such as spontaneous magnetization or another source of valley polarization, would still be required to lift the valley degeneracy. Such spontaneous order could therefore replace the field-induced valley splitting, but it would not by itself reproduce the field-driven suppression of carrier propagation associated with the open orbits in the present mechanism.

Finally, our results highlight the need for theoretical frameworks capable of treating strong correlations in dispersive topological bands. Conventional Hartree-Fock approaches cannot capture fractional states, while exact-diagonalization and DMRG methods face the challenge of identifying an appropriate variational subspace in a regime where neither zero-field minibands nor fully formed Landau levels provide an obvious starting point. More broadly, an important part of the current theoretical understanding of fractional Chern insulators draws on Landau-level-like quantum geometry and vortexability, which connects zero-field band subspaces to the construction of Laughlin-like many-body states. While this paradigm has proven highly successful in flat-band systems, our results suggest that this perspective alone is incomplete in weak-field dispersive regimes where the underlying miniband fermiology remains active. Our findings therefore call for new theoretical tools capable of treating the intertwined roles of fermiology, topology, weak-field quantization, and correlations in this emerging class of dispersive-band quantum phases.

\section*{Methods}
Detailed experimental, fabrication, measurement, semiclassical, tight-binding, and quantum-transport methods are provided in the Supplementary Information.

\section*{Data availability}
The datasets generated during the current study are available from the corresponding authors upon reasonable request. The custom code used for the semiclassical Landau-level predictions is openly available at \url{https://kpm-tools.readthedocs.io/}; no access restrictions apply.

\section*{Acknowledgments}
N.L. thanks Youngju Park for discussions. This work was supported by the Korean NRF through Grant RS-2023-00249414 (N.L.) and Grant NRF-2020R1A5A1016518 (J.J.). We acknowledge computational support from KISTI Grant No. KSC-2022-CRE-0514 and the resources of the Urban Big Data and AI Institute (UBAI) at UOS. S.R. acknowledges Grant PCI2021-122035-2A-2 funded by MCIN/AEI/10.13039/501100011033 and the European Union NextGenerationEU/PRTR and the Departament de Recerca i Universitats de la Generalitat de Catalunya, as well as support from MICIN with European funds--NextGenerationEU (PRTR-C17.I1) and by 2021 SGR 00997, funded by Generalitat de Catalunya. ICN2 is funded by the CERCA Programme/Generalitat de Catalunya and supported by the Severo Ochoa Centres of Excellence programme, Grant CEX2021-001214-S, funded by MCIN/AEI/10.13039/501100011033. C.P.Y.W., W.F., and K.E.J.G. acknowledge funding from the Agency for Science, Technology and Research (Grant No. C230917006). K.E.J.G. acknowledges funding from the National Research Foundation, Singapore, through the National Quantum Office, hosted in A*STAR, under its Centre for Quantum Technologies Funding Initiative (S24Q2d0009). C.P.Y.W. acknowledges A*STAR AME YIRG Grant A2084c0179. J.X.H. and A.A. acknowledge the Singapore Ministry of Education Academic Research Fund Tier 3 Grant MOE-MOET32023-0003, ``Quantum Geometric Advantage,'' and Tier 1 Grant A-8001967-00-00, and the Agency for Science, Technology and Research under its Advanced Manufacturing and Engineering Young Individual Research Grant M23M7c0124. This work was also supported by the Ministry of Education, Singapore, through the Research Centre of Excellence award to the Institute for Functional Intelligent Materials, I-FIM, Project EDUNC-33-18-279-V12 (Z.W. and K.S.N.). K.S.N. acknowledges support from the Royal Society (UK), Grant RSRP R 190000. W.Z. acknowledges A*STAR MTC YIRG Grant M23M7c0122.

\section*{Author contributions}
Z.W. and N.L. conceived the project. Z.W. supervised the experimental work. Z.W. and C.P.Y.W. conducted the transport measurements. Z.W., M.H., J.L., J.H., and L.L. fabricated the transport devices. N.L. supervised the theoretical work. P.M.P.-P. and N.L. performed the simulations. Z.W., P.M.P.-P., H.S., S.R., J.J., K.S.N., and N.L. performed the data analysis. Z.W. and N.L. wrote the manuscript. All authors contributed to data interpretation and manuscript revision and approved the final version.

\section*{Competing interests}
The authors declare no competing interests.

\bibliographystyle{unsrtnat}
\bibliography{all}
\end{document}